# POST-DETECTION SETI PROTOCOLS & METI: THE TIME HAS COME TO REGULATE THEM BOTH

**JOHN GERTZ**
*Zorro Productions, 125 University Avenue, Suite 101, Berkeley, CA 94710, USA.*
Email: jgertz@firsst.org

Regulations governing METI are weak or non-existent. Post-detection SETI protocols are non-binding and too general. Vastly increased SETI capabilities, Chinese involvement in the field, and an intensified effort by METI-ists to initiate radio transmissions to the stars are among reasons cited for urgency in addressing the question of appropriate regulations. Recommendations include regulations at the agency level and laws at the national level as well as international treaties and oversight.

## 1. INTRODUCTION

We know almost, but not quite, nothing about the prevalence and nature of extraterrestrial technologically capable intelligent life (ET). We do know that ET is not obvious. This in itself is a mystery, known variously as Fermi's Paradox or the Great Silence. ET's civilization may be far older than ours, and therefore their technology would likely be far more advanced, allowing ET plenty of time and the means to have colonized the galaxy or to have otherwise made themselves known through fantastical manipulations of their environment detectable from Earth. Many solutions to Fermi's Paradox have been proffered [1]. Some are genuinely bleak, but nevertheless quite plausible. For example, there may be vast numbers of civilizations out there but all are too scared to make themselves seen or heard. What do they know that we do not? In another bleak solution to Fermi's Paradox, there is only one other ET civilization, and it destroys every other life form soon after it commences to emit artificially produced electromagnetic (EM) waves. We might still persist only because we have just recently turned technological, but if this is the correct solution to Fermi's Paradox our annihilation would be imminent. Second, our previous statement that ET *may be* millions or billions of years more advanced than us is too weak. Actually, ET *is* millions or billions of years more advanced than us. We are still in our first century of being capable of sending and receiving interstellar signals. The statistical probability of the ET we might detect also being in its first one hundred years is vanishingly small. We know absolutely nothing about ET's intentions; however, its ability to do us harm, if it so wished, might be absolute. We are only two thousand years more advanced than Rome, yet the best Roman legion could be annihilated by a modern army in mere minutes. Never mind two thousand years ago, Napoleon's armies of barely two hundred years ago would face the same fate.

NASA has within it an Office of Planetary Protection, which is charged with safeguarding the Earth from alien microbes that might otherwise inadvertently attach themselves to our spacecraft returning from missions to planetary bodies. Likewise, it is also charged with doing its best to assure that Earth does not contaminate those bodies with its microbes. NASA, through its Planetary Defense Coordination Office, leads the effort to identify and catalog the orbits of all near earth objects (NEOs) with the potential to collide with Earth. However, NASA has not yet dealt with the risks posed by encounters with ET, whether due to the passive detection of ET's presence, an overt attempt on the part of ET to make contact with Earth, or because we have intentionally initiated contact with them.

SETI and its controversial offshoot, Active SETI or METI [2], are entirely self-regulated activities, though there is a statement of best practices, known as the First Protocol, which we will examine later. There is no generally recognized regulation, law, or treaty forbidding anyone with a transmitter from initiating the transmission of signals to ET or to signal ET in response to a detection of its presence. Kim Jong Un, ISIS, or any other generally recognized rogue actor is entirely free to transmit its particular messages on behalf of all mankind today. In this paper we address the value of global oversight and regulation of communications with extraterrestrial life.

## 2. DRIVERS OF URGENCY

The current author has been an active leader in the field of SETI for more than fifteen years, having served as chairman of the board of the SETI Institute [3] and president of the Foundation for Investment in Research on SETI Science and Technology [4]. As such, he is an enthusiastic supporter of SETI's goal to determine the prevalence and nature of technological civilizations in the universe. However, just as an American supporter of the Second Amendment right to bear arms might also support rational gun control, the author feels that it is critical that the world deal with SETI and METI now through regulations at the agency level, laws at the national level, and treaties internationally. There are five main drivers of urgency.

### 2.1 Enhanced SETI Capabilities

Advances in backend computing power, instrumentation, and detection algorithms make the SETI of today vastly more powerful than the SETI of even a few years ago. Until recently, SETI has been cash starved, being almost entirely reliant upon private philanthropy. Consequently, the entire SETI research team worldwide has rarely exceeded more than the equivalent of twenty full time scientists and engineers.

However, this situation has suddenly changed. In July, 2015, The Breakthrough Foundation announced a $100 million 10-year SETI initiative, called Breakthrough Listen [5, 6]. The combination of greatly enhanced SETI capabilities (for example, ten times wider bandwidth), along with the financial resources to enable those capabilities (for example, much more telescope time), has resulted in a situation where more SETI is currently being accomplished in a single day than has ever been done before in the space of an entire year. Altogether, there has been a concomitant increase in the chances for a detection by at least a factor of 100 over recent times, and yet more orders of magnitude increase over the SETI of past decades.

### 2.2 Increasing Radius of Earth's EM Leakage

Earth has been leaking EM emissions for more than 100 years, meaning that any ET civilization within 100 light years (LY) could, theoretically, have detected our civilization. An estimated 12,000 – 15,000 stars lie within 100 LY, and this so-called *I Love Lucy* spherical radius expands by 1 LY per year. Though Earth's omnidirectional EM leakage is generally quite weak, in that it would require gargantuan receivers to be detectable by ET at interstellar distances [7,8], it is at least hypothetically possible that an advanced civilization within this ever-expanding radius will have detected it, determined that Earth harbors a technological civilization, and may even now be transmitting its response to Earth.

### 2.3 Contact with an ET Probe Could be Imminent.

This author has argued elsewhere that the artificial signal that we first detect may more likely derive from within our own Solar System than from a source outside of it [9]. Just as we have already sent information laden spacecraft into interstellar space, namely, Pioneer, Voyager, and New Horizons, ET may opt to make contact through probes--in effect, sending message in bottles. Probes have several advantages over interstellar radio and optical transmissions. Among them are (a) a much lower cost per byte of transferred information [10]; (b) ET's transmitter and Earth's receivers do not need to exactly line up (each could be cycling through millions of stars, such that the chances of aligning in time would be negligible); (c) security, in that ET need not reveal its location; and (d) a solution to the vexing problem of Drake's "L," in that the transmitting and receiving civilizations need not temporally co-exist. If probes currently reside within the Solar System, then contact with them could be imminent. Imagine for a moment that a probe had landed on an asteroid one billion years ago and burrowed itself into it for protection against radiation and micrometeorite impacts. Further imagine that, knowing Earth to harbor life, once every century or two it resurfaces to scan Earth for artificial EM. It might now suddenly discover that at least one species on Earth has made the transition to technology. This might trigger it to commence broadcasting to Earth from where it is or, alternatively, it might pick itself up and land on the proverbial White House lawn.

### 2.4 China Enters the Field

The Chinese have recently commissioned the largest single dish radio telescope in the world, FAST, and announced that in addition to normative radio astronomy they will use FAST to conduct SETI, partially in collaboration with Breakthrough Listen [11]. Should the Chinese achieve an ET detection separately from Breakthrough Listen, it is unknown whether they would share the news with the rest of mankind or, alternatively, designate it a state secret.

### 2.5 METI-ists Throw Down their Gauntlet

A small, but highly vocal, cadre of scientists has been lobbying hard to commence radio transmissions to putative aliens. This is usually referred to as METI, messaging to extra-terrestrial intelligence, or Active SETI [12, 13, 14, 7]. The present author has recently reviewed the arguments in favor of METI and found them severely wanting. He concluded that METI is unwise, unscientific, unethical, and potentially catastrophic [2].

The author is hardly alone in his opposition to METI. A recent anti-METI petition was signed by such notables as Elon Musk, George Dyson, Dan Werthimer, Geoff Marcy, Paul Davies, David Brin, Michael Michaud, James Benford, among others [15]. Additionally,Stephen Hawking [16] Neil DeGrasse Tyson, Sean Carroll [17], Jarod Diamond [18], as well as, before their demise, John Billingham, Martin Ryle, and Carl Sagan have all registered their opposition to METI.

Nonetheless, a new organization, METI International [19], has recently formed with the intention of engaging in METI as soon as it acquires funding and a willing transmission facility. Its president, Doug Vacoch, has declared that he seeks no broad consensus in initiating transmissions, but rather merely the approval of "peers," presumably a small handful of radio astronomers [20]. METI International's intention, once funded, would be to put in requests for transmission time to transmitting facilities, such as Arecibo, with the hope that the proposal would be peer reviewed on the same basis as any normative astronomical proposal. METI International is not alone in its intentions. Others have also either already conducted limited METI or have announced their intention to do so [21,22, 23, 24, 25]. Peer review, presumably by radio astronomers, is woefully inadequate for the occasion. In the words of retired State Department diplomat, Michael Michaud, "… [SETI] is science and is properly the domain of scientists. Deliberately calling attention to ourselves is not science, but policy" [26].

## 3. WHAT TYPE OF PROBLEMS MIGHT REGULATIONS, LAWS, AND TREATIES SOLVE?

### 3.1 Post-Detection Dissemination of Information

In the event of a detection, does the world have a right to know the fact of a detection, the coordinates of a detection, and the content, if any, of a message? The current ethos, as reflected in the First Protocol [27, 28], is a resounding "yes" to all of the above. The current author is personally acquainted with most major Western SETI scientists and is confident that maximal transparency is their intention. However, there are no assurances that scientists would not feel compelled to first get permission from their respective internal security services, and that such permission might not be forthcoming. In fact, the author has heard from serious U.S. SETI researchers that they are convinced that "men in black suits" will appear at their laboratory door the moment a detection is confirmed. In effect, Russia, China, or even the United States might stamp the very fact of a detection, much less the coordinates and the content, as top secret. In the U.S., in the absence of laws to the contrary, free speech considerations might prevail. However, this is not necessarily true about detections made in countries without strong safeguards governing the free dissemination of scientific findings. Moreover, there is room to question whether scientists should readily make available to the public the coordinates of the message. Perhaps that element should remain secret until such time as mankind has determined whether and how to reply. Otherwise, some party in possession of a transmitter might broadcast its own parochial response, without any broad consensus to do so. The author concedes that, at least among Western democracies, hiding the coordinates would be difficult, given that numerous telescopes would (and should) be enlisted in the follow up confirmatory observations.

Leaving aside for a moment the question of feasibility, let us ask the question of whether total, unfiltered, and near real time transparency, as envisioned by most Western SETI scientists, is advisable. In the author's opinion, there is no valid reason to hide the fact of an ET detection from the public. However, the immediate release of the coordinates of a transmission begs an unauthorized and premature response. Religious groups might send their parochial messages, while Kim Jong Un might send his. Most SETI scientists believe that our current telescopes are not sensitive enough to detect a radio message that might be embedded within ET's carrier wave. Moreover, the consensus view is that it might take many years, if at all, to decode any message, even it were detectable. We still barely understand anything of the Mayan writing system, even though it was devised by members of our own species. Nor do we understand almost anything of dolphin, squid, or other communications of non-human species. Perhaps mankind should not rush to respond to a message it cannot even decipher. However, as a hypothetical, let us consider a message from an ET probe located within our own Solar System [9]. Such a message might be loud enough to be decipherable, and could even be in perfect English or some other terrestrial language if ET had been studying Earth's transmissions over the past century. Further imagine that the message is not along the lines of "Peace be with you…," but more like "We have been watching your TV news broadcasts and see that you are a wretched and violent species, so we therefore give the gift of a recipe for a world destroying self-replicating toxin. Enjoy." It is an outlandish hypothetical, to be sure, but it indicates the possible need for some system of vetting before release to the general public.

If there is a consensus that not necessarily all data should be made available to the general public, then the question becomes how, in the age of social media, a partial information embargo might be enforced. Though not simple or susceptible to an iron-clad guarantee, there are three means of protecting the data. First, there needs to be ethical endorsement by SETI scientists. Smallpox and other pathogens are not released from the few laboratories authorized to handle them by virtue of the clear understanding of the scientists involved of the attendant dangers. Second, there needs to be widely agreed procedures for correctly handling the data. For example, the data might be given over to, and be in the sole possession of, a standing committee of distinguished scientists and policymakers appointed by, and reporting directly to, the UN Security Council. That committee would be authorized to farm out sub-sections of the data to appropriate researchers. For example, linguists might be charged with studying the message, but might never be given the coordinates. Coordinates might be given only to a handful of approved astronomers for follow up observations. Third, there might be criminal laws put into place, similar to existing laws against the release of top secret government. Criminal penalties and science are not normally mentioned in the same breath. So to be clear, all SETI up to the moment of signal or artefact detection should be properly viewed as pure science and free from any regulation or coercion. Actions undertaken post-detection, apart from confirmatory observations, are not science, but matters of vital public policy. This also pertains to pre-detection self-initiated signaling by METI-ists. In practice, it is inconceivable that SETI scientists or METI-ists would ever actually be jailed for unauthorized signalling, partly because they are assumed to be law-abiding citizens, but also because neither the directors of transmitting facilities, nor the necessary support staffs, would likely become complicit in a criminal act.

### 3.2 Who Owns an ET Probe?

Our Solar System could be populated with large numbers of ET probes that have been sent to passively observe Earth over millions or billions of years. SETI scientists have only recently accepted this paradigm and are now beginning to search for them. However, in the meantime, an ET probe might land on the White House lawn. But it could just as easily land on Kim Jong Un's lawn or in front of the Taj Mahal. It could also be discovered by accident by a privately owned mining expedition on an asteroid, or it could be broadcasting to us and detected by a radio telescope, and then be retrieved by a spacecraft owned by whichever country or private company, such as Space X, might get there first. Does the first country, company, or even billionaire acting as an individual that retrieves a probe own it? Who gets to examine it first? Should it be examined *in situ* where it is discovered, say, on an asteroid, or brought back to Earth, where it could presumably undergo more thorough examination? After all, the probe might harbor infectious lifeforms, or in some other fashion be an intentional Trojan horse. By treaty, should the probe belong to all mankind and be administered scientifically by the UN Security Council, or by a pre-designated research institution? How is the probe to be treated? The probe may not be just a piece of machinery, but also ET's artificial life form and ambassador, not dissimilar to C-3PO of Star Wars. Should it be given diplomatic immunity and every courtesy normally afforded an ambassador? Who is to make all of these decisions?

### 3.3 Proscribing METI

In this author's view, METI, as operantly defined as the intentional transmission to interstellar targets of EM signals of a strength greater than Earth's omni-directional leakage, should be considered as the reckless endangerment of all mankind, and be absolutely proscribed with criminal consequences, presumably as exercised at the national level, or as administered through the International Court of Justice in The Hague. The one thing that our theories tell us about ET—and this is a virtual certainty—is that ET is far, far more advanced technologically than we. Imagine what even a great thinker and tinkerer like Benjamin Franklin would think if he returned today to find smartphones, moon landings, and heart transplants. ET's technology might similarly seem indistinguishable from magic to us. We cannot know today whether ET might intend to harm us, benefit us, proselytize, or simply trade knowledge. However, if their intent be malign, they might have the ability to destroy us from afar with a single projectile filled with a self-replicating toxin or nano-grey-goo. We simply do not know today anything of ET's intent. SETI seeks to study the universe in order to answer the question of whether ET resides or lurks out there at all, and if so, to learn as much as possible about it. However, the decision to blindly shout out to ET, inviting possible catastrophe, is not one that a few lonely radio astronomers should have a right to make on behalf of all of mankind. The idea of jailing scientists may seem harsh (and, as stated earlier, is in practice highly unlikely). However, though wearing pseudo-scientific hats, METI-ists are actually engaged in unauthorized and frankly unethical diplomacy. It is not for scientific excellence in biology that someone who cultures and releases anthrax would be prosecuted as a criminal.

### 3.4 Proscribing Unintentional METI

When used as a transmitter, Arecibo is the most powerful radar on Earth, whose beam might be readily detectable at interstellar distances. Fortunately, the beam is narrow, its use as a radar has been rare, and the targets have not been nearby stars, but, usually asteroids. Other such powerful transmitters include Goldstone in California and Evpatoria in Ukraine, and transmitters would not be difficult to add to FAST or some other radio telescopes. Additionally, some military and space agency radars are also quite powerful, though far less so than Arecibo or other radio telescopes used as transmitters. Treaties might specify the legitimate purposes for these radars, and each transmission might be subject to review committees especially informed by a prohibition against METI. Thus, for example, the use of radars above a specified intensity thresholds might be prohibited when the target of interest occults any star within, say, 1000 LY (covering the local thickness of the Milky Way), or when transiting the plane of the Milky Way.

## 4. CURRENT AGREEMENTS

There are two putative agreements that govern SETI/METI; (a) The Declaration of Principles Concerning Activities Following the Detection of Extraterrestrial Intelligence adopted by the International Academy of Astronautics in 1989, and as revised in 2010, usually referred to simply as The First Protocol [27, 28]; and (b) The Treaty on Principles Governing The Activities of States in the Exploration and Use of Outer Space including the Moon and Other Celestial Bodies as ratified by the UN in 1966. There are currently 104 signatory nations. This treaty deals with space exploration, a nation's responsibilities for its actions in outer space, nuclear weapons in space, and other facets of space law [29]. There is also the Moon Agreement adopted by the U.N. General Assembly in 1979, but never ratified by the U.S. Senate [30].

### 4.1 First Protocol

The nine Principles of the First Protocol for present purposes can be distilled into three broad concepts: (a) SETI shall be conducted transparently; (b) a detection should be followed by rigorous confirmatory procedures and follow up observations; and (c) everyone must refrain from transmitting a response without authorization from a broadly representative body, such as the Security Council of the United Nations [31]. There are several problems with the First Protocol:

- These principles are not recognized as law by any government. Though it is endorsed informally by most current SETI scientists as a system of best practices, the First Protocol is merely the work product of a standing SETI committee of the International Academy of Aeronautics (IAA).
- METI-ists generally agree to abide by the First Protocol so far as it pertains to post-detection replies, but only that. Principle #8 of the First Protocol states "In the case of the confirmed detection of a signal, signatories to this declaration will not respond without first seeking guidance and consent of a broadly representative international body, such as the United Nations." Logically, if unauthorized signaling after a detection is proscribed, METIists should not be able to claim that self-initiated transmissions prior to a detection are permissible. However, without any conceivable logic, they deny that the First Protocol applies to pre-detection transmissions. In other words, after First Contact, they are willing to "consult," but before First Contact, they grant themselves full permission to transmit freely.
- The First Protocol is silent on all other matters, including, for example, the question of who owns an ET artifact, such as a probe.
- Chinese SETI scientists have not pledged fealty to the First Protocol, nor has the Chinese government.
- The First Protocol may not stand up to the excitement and pressures of the moment. In the age of social media, is it even likely that the Protocol will be adhered to? Would the detection be announced to the world only after very careful steps have been taken to confirm the authenticity of the detection, and then only by the discovering team in a press conference, as envisioned by the First Protocol? Or would the information leak to the press at a much earlier stage in the process? There is no current law against posting information of the detection on Twitter or Facebook. It merely violates a statement of best practices, with which an overly enthusiastic undergraduate intern may be unfamiliar as he makes the post [32].
- There is no vetting process in place that might filter information before it is released to the public. Although drafted by veteran State Department diplomat, Michael Michaud, there was no higher input from the State Department, Defense Department, security services, Executive Branch, National Academy of Sciences, or world leaders. Would a higher level treaty agree with the principle of total transparency?

In sum, the First Protocol is wholly inadequate in the face of the urgencies listed above in Section 2 of this paper.

### 4.2 Treaty on Principles Governing Activities of States in the Exploration and Use of Outer Space

Treaties, when ratified by the U.S. Senate theoretically enjoy the full force of law. It can be reasonably inferred that this treaty already governs SETI and METI, but only generally.

#### *4.2.1 Article I Supports SETI*

Article I states that there shall be freedom of scientific investigation in Outer Space, hence broadly condoning SETI, which seeks to investigate the prevalence and nature of intelligent life in the universe. It further holds that all scientific results shall be for the benefit of all mankind.

#### *4.2.2 Article IX Proscribes METI*

Article IX states that no experiment may be conducted with the potential to harm Earth may be conducted without "appropriate international consultations."

#### *4.2.3 Treaty Drawbacks*

METI-ists do not recognize its authority or applicability. Even the most fervent METI-ist agrees that METI has an element of risk. However, on the basis of absolutely no data whatsoever, they regard the risk as small and/or worth the potential gain. This author argues elsewhere [2] that it is impossible to measure the risk today along any parameter whatsoever. All we can say is that there is risk, but we cannot know whether the risk of a bad outcome is highly probable or not; or whether the nature of that risk is catastrophic or mild. Because everyone admits that there is a risk of harm to planet Earth, METI should therefore already be proscribed under the terms of this treaty. Yet it has not been. METI-ists simply ignore this feature of the treaty, and neither the United States nor any other country has as of yet chosen to enforce the clear meaning of the treaty within its borders.

The MET-ists do have an argument insofar as the treaty does not explicitly address activities related to space that are ground based, such as METI transmissions.

There is no enforcement mechanism or stated penalties. This was made apparent when the treaty was grossly violated by the Chinese in 2007 when they intentionally destroyed one of their own weather satellites with a missile, creating thousands of pieces of space junk, jeopardizing all low earth orbit satellites. China's reckless action was met with broad condemnation, but no punitive consequences.

### 4.3 Agreement Governing Activities of States on the Moon and Other Celestial Bodies

This treaty never came into full effect, failing to be ratified by the U.S. Congress, or indeed by other major countries, such as China and Russia. Nevertheless, it is instructive. Article 5.3 of this treaty reads: "In carrying out activities under this Agreement, States Parties shall promptly inform the Secretary-General, as well as the public and the international scientific community, of any phenomena they discover in outer space, including the moon, which could endanger human life and health, as well as of any indication of organic life." [33] On its face, this would seem to lend some support to the First Protocol in calling for complete transparency. This may be true of the spirit of the Agreement, but not necessarily of the letter. Although Article 5.3 refers to phenomena discovered in "outer space," Article 1 indicates that the subject of the Agreement is limited to the moon and other celestial bodies within the Solar System. Hence, "outer space" can be regarded as bordered by the Oort cloud for the purposes of Article 5.3. An interstellar detection need not be revealed. Nor would a detection need be revealed if not known to be a threat nor known to be organic. A detection might be interpreted as coming from a non-organic source, such as AI.

## 5. RECOMMENDATIONS

The author would like to suggest a set of complementary recommendations that might be mutually reinforcing, allowing the considered problems to be attacked from various policy angles at once.

### 5.1 Seek Injunctive Relief Against METI in Courts of Law and Department of State for ITAR Violations

Pursuant to Article IX of the Space Treaty, METI would arguably be illegal. Consequently, on its face, it would seem that organizations or individuals attempting to conduct METI might be enjoined in a court of law. Actions of declaratory relief might therefore be filed against managing agencies of transmitting facilities such as Arecibo. However, such actions could be fraught with problems. It is not certain whether courts would rule that METI violates the treaty. Even before that stage of the argument, they must determine whether a private citizen or organization has any private rights under the treaty. The Supreme Court, in its 1998 ruling in *Medellin v. Texas,* threw great doubt on whether individuals have standing with regard to treaties that deal with issues of human rights, as opposed to issues related to property rights. Since then, federal courts have been wont to rule that treaties were not meant to benefit individuals. Additionally, in 2001, the Fifth Circuit Court, in *McKesson v. Islamic Republic of Iran*, drawing on *Medellin,* found that the Treaty of Amity created rights, but was silent as to remedies. The Space Treaty is also silent as to what actions might be taken to enforce it, and Congress never passed enabling legislation. Nonetheless, suits might be filed, as METI does raise issues not previously placed before the courts, namely, the rights of all mankind as allegedly imperiled by a few radio astronomers [34].

However, suits by individuals or organizations seeking injunctive relief against METI would not be necessary if governmental bodies, such as the White House Office of Science and Technology, FCC, NSF, NASA or, the Departments of State and/or Defense declared METI as a violation of our international obligations under the Space Treaty. Then the shoe would be on the other foot, and it would be the METI-ists who would have to sue the government to seek relief.

Alternatively, an action might be filed in the International Court of Justice, where it might have a better likelihood of success. A decision against METI might not be directly enforceable in the United States, but would probably be honored voluntarily by agencies with authority over transmitters.

METI would seem to violate International Traffic in Arm Regulations (ITAR). Information and technology, not in the public domain, which could be militarily helpful to a foreign entity, is prohibited from transfer without a special waiver from the Department of State. METI-ists have indicated that they are prepared to beam up the entire content of the Internet to ET, the ultimate foreign entity, essentially giving ET everything they need to know about our level of technological advancement and hence our defense capabilities. U.S. METI-ists should be compelled to seek a State Department waiver before proceeding, or face stiff penalties as already prescribed under the law. METI-ists might plausibly claim that the information that they would transmit to ET is in the public domain, and hence not covered by ITAR. However, though the information may be in Earth's public domain, it is entirely unknown to ET.

### 5.2 National Academy of Sciences (NAS) Study

The advantages of an NAS study are that its recommendations are often viewed as authoritative by various arms of the government. A positive NAS finding towards SETI might eventually result in funding by NASA, NSF, or directly through Congress. An NAS study also has great weight beyond the borders of the U.S. However, the disadvantage is that the empaneled members of the NAS study group would likely be comprised mostly of scientists, leaving out experts in relevant fields beyond the hard sciences. NAS studies are generally ordered and, crucially, funded by Congress or another arm of government, though other entities may also commission studies. Fortunately, many SETI scientists and known SETI sympathizers are NAS members, and SETI also has some friends in Congress. SETI proponents might be organized to request a formal NAS review, while a foundation or wealthy individual might agree to fund it, if Congress or an agency of government, such as NASA, does not step forward. Let the chips fall where they may, but this author is of the opinion that such an NAS study would make the following two baseline recommendations:

- SETI is a serious and cost effective scientific discipline aimed at determining the prevalence and nature of technological intelligence in the universe. The fact that SETI is serious hardly requires defense. What is often less appreciated is that it is much cheaper to conduct SETI experiments than to determine whether there was ever a separate genesis of life in our Solar System on such planets and moons as Mars, Europa, Enceladus, or Titan, or on exo-planets. NASA currently spends next to nothing on the former, and billions of dollars on the latter. It will cost many billions of dollars in space missions to make a determination as to whether there is another inhabited body in our Solar System or to launch space telescopes capable of taking spectra of the atmospheres of exo-planets, whereas the current worldwide annual budget for SETI is approximately only $10 Million. In other words, for much less than a penny on the dollar one can leapfrog the entire question of life on other bodies within our own Solar System or on exo-planets, and address the more (or at least "as") interesting question of technologically competent intelligent life.

  Further, if participation in the consideration of SETI can be broadened to include defense officials and public policymakers, the implications for national and world defense may come to be fully appreciated. We are currently utterly blind to ET's prevalence, capabilities and intentions. If there are very bad actors out there in the cosmos, the only way that we can learn anything about them is through a well-funded SETI program. It is the same logic that impels us to search for and describe the orbital characteristics of NEOs. Whereas SETI scientists have sought, mostly in vain (at least in recent times), for NASA or NSF funding, it may be the Defense Department that should be considered the more logical funding source.

  This author would be shocked if NAS did not full-throatedly endorse SETI as a wise use of public science dollars.

- METI is not science, but diplomacy, which, as such, should not be initiated without the expressed approval of the UN Security Council.

### 5.3 A SETI Congress and the Asilomar Process

Asilomar is a wooded, seaside conference center outside of Monterey, California. It is famous for a 1975 conference on the bioethics of recombinant DNA. Representation was worldwide, and included not just multiple scientific disciplines, but also legal scholars, members of government and the press. The meeting's work product was a set of guidelines that have largely governed recombinant DNA research ever since.

The advantages of the "Asilomar process" over an NAS study are that (a) it would be an international effort; and (b) it could bring in experts from a wide array of fields beyond astronomy among the sciences and beyond the hard sciences altogether, including historians, economists, global security experts, ethicists, diplomats and so forth.

The main disadvantage is that its work product would not have any force of law, or immediately result in loosened government purse strings. An NAS study commissioned by Congress might have more influence on Congress with respect to formulating relevant legislation, including appropriations. As with the First Protocol, its recommendations would rely on voluntary compliance. This has worked pretty well in the field of recombinant DNA, but may not be effective for SETI and METI. Modern medical research regularly works within the constraints of ethical review panels. The very concept of experimental ethics is foreign to astronomers, not, of course, because they are unethical people, but because their science is normally observational, rather than manipulative, and they study their subjects literally from afar.

METI is unique in astronomy, in that it is manipulative, seeking to influence the universe rather than just study it. Additionally, this author has argued that METI has more in common with a religious cult than with science [2]. METI-ists implicitly believe that ET is omniscient, all good, and omnipotent. Omniscient, because ET knows we are here, even though Earth's omni-directional EM leakage is very weak; all good, because ET is only interested in our welfare and has no malign intent; and omnipotent, because even though METI-ists make no provisions at all for the receipt of return messages, they believe that ET will respond, somehow, some way, perhaps by taking over terrestrial TV broadcasts or YouTube. Unfortunately, the most fanatic METI-ists are unlikely to be deterred by any broad-based consensus of best practices, especially one that lacks teeth.

Such an Asilomar conference devoted to SETI and METI would have to be funded at its inception by a foundation or by a single visionary wealthy individual. The envisioned conference should be organized in such a fashion as to best assure a quality end product, that being a statement of principles and a set of workable rules and procedures, governing METI and postdetection SETI protocols. As such, it might better be called a "congress" than a "conference."

The envisioned SETI Congress, of course, need not be held at Asilomar, and might be sponsored and administered by any number of responsible parties, acting alone or in some concerted fashion, such as, one or more major universities or relevant institutes that they house, the Breakthrough Foundation, the Royal Academy, the SETI Institute, the United Nations, and so forth. The Congress' crucial conditions should be that it be (a) ecumenical, in that delegates represent highly variegated expertise; (b) distinguished, in that its members be leading authorities in their disciplines; and (c) thorough, in that it take the amount of time and effort in its preparation and execution to best assure a quality work product with enduring significance and consequence.

### 5.4 Amend the UN Treaty on Space, as Allowed Under Article XV

The current author would recommend three explicit amendments:

- **Prohibit METI:** METI, as operantly defined as the intentional signaling to extra-Solar targets Earth's presence as a technological civilization at a signal intensity greater than that of a terrestrial TV broadcasting tower (i.e., the "I Love Lucy" power level that has long since left Earth), should be prohibited.
- **Apply best practices limiting inadvertent METI:** Powerful non-METI radar transmissions towards satellites, asteroids and similar targets would follow best practices of desisting during the time when the target transits the plane of the Milky Way or occults a background star, so that their beams do not inadvertently illuminate those stars.

- Create and maintain a standing committee governing SETI and METI whose members shall be comprised of scientists, policy makers and other relevant parties reporting directly to the UN Security Council.

.

### 5.5 Treaties Between or Among U.S., Russia, Ukraine, China, India, South Africa, Australia and Other Potential Transmitters

There are actually only a handful of countries with serious SETI and METI capabilities. In the absence of a broader international treaty, smaller multinational treaties might suffice.

### 5.6 National Laws Proscribing METI

National legislative bodies could pass laws against METI, explicitly citing it as a violation of the Space Treaty, and explicitly enumerating criminal penalties. No other action could be stronger than this. In the United States, a METI-ist's only recourse would be a court ruling, overturning the legislation, presumably on free speech grounds.

Such legislation might also specify that the fact of a detection can be made public by the scientists who made it, but the coordinates and content should go only to the President of the U.S., who, with his national security team, can determine from there whether and what information to release to the public. The advantage of national laws over international treaties is that criminal penalties are normally administered at the national level. So even in the presence of treaties, national laws can act as an enforcement buttress.

### 5.7 FCC, NSF and NASA Regulations Against METI

Relevant agencies of the Federal government, even in the absence of international agreements, can adopt rules proscribing METI and limiting unintentional METI along the lines already suggested. The FCC could ban METI transmissions; NSF could prevent facilities managed by it, such as Arecibo, to be used for the purposes of METI. NASA's Office for Planetary Protection could also weigh in. This office has the authority to quarantine sample return missions so that they do not inadvertently contaminate our planet with unknown biohazards from outer space. It also oversees NASA's efforts to avoid contamination of Solar System bodies by our landing craft. Clearly this office has an interest in the potential harm to planet Earth that METI might cause.

### 5.8 Create a Standing Committee of Experts

As mentioned, there is a SETI standing committee under the IAA, however, it is woefully inadequate, being comprised, in the most part, by astronomers, and it has no enforcement authority whatsoever. The committee this author envisions would include experts in military and civil emergency planning, intelligence, diplomacy, economics, and scientific fields outside of astronomy, such as linguistics, evolutionary biology, computer science, and others. The committee would study SETI and METI, issuing a report. In effect, it would replicate the envisioned NAS study, except that it would be an international effort and reflect broader than hard science only input. The standing committee would be charged with regulating radar transmissions and guarding against intentional METI. Finally, it would manage post-detection confirmations and reactions, such as advising if and how to respond. The author's preference is that the envisioned standing committee be international, and organized by and reporting to the UN Security Council. Failing that, the committee would be international, and organized by a coalition of those nations in the possession of receivers and transmitters relevant to SETI and METI. Failing that, the standing committee would be organized by an agency of the U.S. government, and ultimately report to the President. Failing that, the committee would be organized by and report to NASA's Office of Planetary Protection. Failing that, the standing committee would be constituted by the SETI Congress envisioned in Section 5.3 above and, in effect, be left behind after the Congress had finished its main report.

## 6. CONCLUSIONS

The author resides in California, where we are taught to prepare for a catastrophic earthquake ("the Big One"). We do not know where and when it will hit, but we know it is coming. Similarly, failing action to divert them, Earth will be hit by asteroids of a size that might imperil. We do not yet know the specifics, other than there will be direct hits. The study of NEOs has been ongoing for some years in an effort to identify the where and the when of the inevitable. We cannot say today when first detection of an alien civilization will take place. However, with the ever quickening pace of our SETI efforts, it is generally believed among most of those involved in the search that if ET is out there, it may not be long now before a detection. The analogy between an ET detection and the Big One or asteroid collisions is not perfect. Whereas the latter are a certainty, we cannot promise that a detection will ever happen, much less the timing. *What we can say is that if it is going to happen*, it may happen sooner rather than later.

The present paper was written in an attempt to persuade that serious planning for the day of and the days following a detection begin now. The world must agree upon, and bring into effect, some rigorously defined and enforced procedures, where currently there are none. Far better that those policies and procedures be the product of thoughtful deliberations and be put in place before a detection is made than they be improvised in great haste immediately after the fact.

METI is not SETI. It has been allowed to proceed at its own pace and without any oversight simply because the general public and policy makers have yet to grapple with the significance and seriousness of its hubris. The world must awaken to its potential consequences.

## REFERENCES

1. S. Webb, "*Where is Everybody?: Fifty Solutions to the Fermi Paradox and the Problem of Extraterrestrial Life*", Springer, New York, 2002.
2. J. Gertz, "Reviewing METI: A Critical Analysis of the Arguments", *JBIS*, **69**, pp.31-36, 2016.
3. SETI Institute Homepage, http://www.seti.org/. (Last Accessed 28 November 2016)
4. FIRSST Homepage, http://firsst.org/landing/. (Last Accessed 28 November 2016)
5. Breakthrough Initiatives, "Yuri Milner and Stephen Hawking Announce $100 Million Breakthrough Initiative to Dramatically Accelerate Search for Intelligent Life in the Universe", http://www.breakthroughinitiatives.org/News/1. (Last Accessed 10th February 2016)

6. H. Isaacson, A. Siemion, G. Marcy, M. Lebofsky, D. Price, D. MacMahon, S. Croft, D. DeBoer, J. Hickish, D. Werthimer, G. Hellbourg and J. Emilio Enriquez, "The Breakthrough Listen Search for Intelligent Life: Target Selection of Nearby Stars and Galaxies", *PASP*, to be published in 2017.
7. J. Billingham and J. Benford, "Costs and Difficulties of Interstellar 'Messaging' and the Need for International Debate on Potential Risks", *JBIS*, **67**, pp.17-23, 2011.
8. A. Penny, "LOFAR SETI Sources From Terrestrial Analogues", ASTRON Presentation, 2008
9. J. Gertz "E.T. Probes: Looking Here As Well As There", *JBIS*, **69**, pp. 88-91, 2016.
10. C. Rose and G. Wright, "Inscribed Matter as an Energy-Efficient Means of Communication with an Extraterrestrial Civilization", *Nature*, **431**, pp.47-49, 2004.
11. A. Siemion, "National Astronomical Observatories of China and Breakthrough Initiatives Collaboration", https://breakthroughinitiatives.org/News/6. (Last Accessed 28th November 2016).
12. S.J. Dick, "Should We Message ET?", http://www.setiinternational.org/blog/should-we-message-et. (Last Accessed 9th February 2016)
13. D. Brin, "Shouting at the Cosmos", http://lifeboat.com/ex/shouting.at.the.cosmos. (Last Accessed 10th February 2016)
14. S. Shostak, "Are Transmissions to Space Dangerous?, *Int. J. of Astrobiol.*, **12**, pp.17-20, 2013.
15. "Regarding Messaging to Extraterrestrial Intelligence (METI)/Active Searches for Extraterrestrial Intelligence (Active SETI)", http://setiathome.berkeley.edu/meti_statement_0.html. (Last Accessed 9th February 2016)
16. C. Moskowitz, "If Aliens Exist, They May Come to Get Us, Stephen Hawking Says", http://www.space.com/8288-aliens-exist-stephenhawking.html. (Last Accessed 10th February 2016)
17. SETI League, "Declaration of Principles Concerning Activities Following the Detection of Extraterrestrial Intelligence", http://www.setileague.org/general/protocol.htm. (Last Accessed 9th February 2016)
18. J. Diamond, "To whom it may concern", *NY Times Magazine*, **5**, pp.68-71, 1999.
19. Mission Statement, http://meti.org/. (Last Accessed 10th February 2016)
20. D. Vakoch, "In Defense of METI", *Nature Physics*, **12**, pg. 890, 2016.
21. Diamond Sky Productions, "Message to the Milky Way", http://diamondskyproductions.com/recent/index.php#mmw. (Last Accessed 18th February 2016)
22. "Artistic Space Odyssey to Broadcast People's Messages to the Stars", Phys.org, http://phys.org/news/2016-02-artistic-space-odyssey-people-messages.html. (Last Accessed 18th February 2016)
23. A. Zaitsev, "Sending and Searching for Interstellar Messages," *Acta Astronautica*, **63**, pp.614-617, 2008.
24. A. Zaitsev, "Detection Probability of Terrestrial Radio Signals by a Hostile Super-Civilization", arXiv:0804.2754. (Last Accessed 19th February 2016)
25. J. Haqq-Misra, M.W. Busch, S.M. Som and S.D. Baum, "The Benefits and Harm of Transmitting into Space," *Space Policy*, **29**, pp.40-48, 2013.
26. M. Michaud, "*Contact with Alien Civilizations*", Copernicus Books, New York, 2007.
27. "Protocols For an ETI Signal Detection," SETI Institute, http://www.seti.org/post-detection.html. (Last Accessed 28th November 2016)
28. "Declaration of Principles Concerning the Conduct of the Search for Extraterrestrial Intelligence", SETI Permanent Study Group of the International Acadamy of Astronautics, 9.30.2010.
29. U.N. General Assembly Resolution 2222 (XXI), "Treaty on Principles Governing the Activities of States in the Exploration and Use of Outer Space, including the Moon and Other Celestial Bodies", *1499th plenary meeting, 19 December 1966.*
30. P. Ney, "Extraterrestrial Intelligence Contact Treaty?", *JBIS*, **38**, pp.521-522, 1985.
31. SETI League, "Declaration of Principles Concerning Activities Following the Detection of Extraterrestrial Intelligence", http://www.setileague.org/general/protocol.htm. (Last Accessed 9th February 2016)
32. S. Shostak, "Media Reaction to a SETI Success", *Acta Astronautica*, **41**, pp.623-627, 1997.
33. U.N. General Assembly Resolution 34/86, "Agreement Governing the Activities of States on the Moon and Other Celestial Bodies", *89th plenary meeting*, 5 December 1979.
34. O. Hathaway, S. McElroy and S. Solow, "*International Law at Home: Enforcing Treaties in U.S. Courts*", Yale Law School Legal Scholarship Repository, 2012.

* * *